\newcommand{\pla}[3]{Phys. Lett. {\bf #1{A}}, #3 (#2)}

\newcommand{\jpa}[3]{J. Phys. A:Math. Gen. {\bf #1}, #3 (#2)}

\newcommand{\jsp}[3]{J. Stat. Phys. {\bf #1}, #3 (#2)}

\newcommand{\Lm}{\Lambda}
\newcommand{\ra}{\rightarrow}

\documentstyle[aps,eclepsf,preprint]{revtex}

\begin{document}
\draft
\title{Test of Guttmann and Enting's conjecture in the eight-vertex model}
\author{Hiroshi Tsukahara and Takeo Inami}
\address{Department of Physics, Chuo University, Kasuga, Bunkyo-ku, Tokyo 112, 
Japan
}
\date{\today}
\maketitle

\begin{abstract}
 We investigate the analyticity property of the partially resummed series 
expansion(PRSE) of the partition function for the eight-vertex model. 
Developing a graphical technique, we have obtained a first few terms of the 
PRSE and found that these terms have a pole only at one point in the complex 
plane of the coupling constant.
This result supports the conjecture proposed by Guttmann and Enting concerning
the ``solvability'' in statistical mechanical lattice models.
\end{abstract}

\narrowtext

 Recently Guttmann and Enting introduced a new point of view into the study 
of the solvability of lattice models in the statistical mechanics, including 
models for combinatorial problems\cite{guttmann-enting:1996}. 
 Their approach is based on a study of the connection between the solvability
of lattice models by use of the inversion relation and the analyticity 
properties of their associated series expansion, which is called the partially
resummed series expansion(PRSE).
 For exactly solvable models, the inversion relation for the partition function
was derived using the star-triangle relation\cite{stroganov:1979,baxter:1982}. 
Subsequently, it was shown to hold true for the partition function of (so far)
unsolved models and other physical quantities as well
\cite{jaekel-maillard:1982-1983}.
Thus the inversion relation can be used for clarifying the difference between 
exactly solved quantities and unsolved ones in statistical lattice models.

 The importance of the analyticity properties of the PRSE was first pointed 
out by Baxter\cite{baxter:1980}. He considered the PRSE for the reduced free 
energy $\ln\Lm(t_1,t_2)$ of the zero-field anisotropic Ising model on the 
square lattice,
$\ln\Lm(t_1,t_2) = \sum R_n(t_1^2)t_2^{2n}$, and showed that the functions
$R_n(t_1^2)$ can be determined recursively from the inversion relation,
$\ln\Lm(t_1,t_2) + \ln\Lm(t_1^{-1},-t_2) = \ln(1-t_2^2)$, 
and the symmetry relation $\Lm(t_1,t_2) = \Lm(t_2,t_1)$,
provided $R_n(t_1^2)$ have singularities only at a single point in 
the $t_1^2$-complex plane.
 This realization led to a series of investigations on the partial sums 
$R_n(t_1^2)$ for unsolved quantities such as the partition 
function of the 2D Ising model in a field\cite{baxter:1980}, non-critical
2D Potts model\cite{jaekel-maillard:1982-1983}, zero-field 3D Ising model
\cite{hansel-et.al.:1987} or the susceptibility of the 2D Ising model
\cite{guttmann-enting:1996}.
The results for these quantities indicate that infinitely many poles appear in
partial sums $R_n$ in the limit $n \ra \infty$ in clear contrast with the
solvable case.

 The new approach of Guttmann and Enting to the solvability of lattice models
is as follows.
They defined a given quantity to be solvable if its solution can be expressed 
with a D-finite function\cite{stanley:1980} {\em as a function of the coupling
constant}.
 D-finite functions are defined as functions which satisfy a linear
ordinary differential equation of finite order with polynomial coefficients. 
The definition of the solvability was extended so that solvable quantities
do not necessarily have a closed form solution. However, it is restrictive
in the sense that those quantities which are expressible with the solution of 
a nonlinear differential equation are classified as ``unsolvable'' ones.

 Guttmann and Enting further put forward a criterion which determines whether 
a given quantity is solvable or not in the sense of D-finiteness.
First, generalize the model such that it possesses two (or more) anisotropic
coupling constants $t_1$ and $t_2$. Let the wanted quantity be $f(t_1,t_2)$. 
One can calculate exact coefficients of the series expansion of $f$,
$f(t_1,t_2) = \displaystyle \sum_{mn} a_{mn} t_1^m t_2^n$
(to a considerably high order with sophisticated techniques).
Then, performing the infinite sum with respect to, say, $t_1$ first, 
one obtains the PRSE,
$f(t_1,t_2) = \displaystyle \sum_n R_n(t_1) t_2^2$.
Let us assume that the poles in $R_n(t_1)$ all lie on the unit circle in the 
$t_1$-complex plane (this is not essential, however).
Their conjecture on the criterion is stated as follows:
(i) If the number of poles of $R_n$ remains finite as $n \ra \infty$, then 
$f(t_1,t_2)$ is a D-finite function of $t_1$ with $t_2$ being fixed.
(ii) If there appear infinitely many poles densely distributed along the unit 
circle as $n \ra \infty$, then $f(t_1,t_2)$ cannot be expressed in terms of 
D-finite functions only.

 It is illuminating to look at the spontaneous magnetization $M$ of the 2D
Ising model as an example of the case (i). The closed form solution is given 
as $M(u,v) = (1-16uv/(1-u)^2(1-v)^2)^{1/8}$ in the low temperature phase,
where $u = (1-t_1)/(1+t_1)$ and $v =  (1-t_2)/(1+t_2)$.
One can calculate a first few terms of the Taylor expansion 
$M(u,v) = \sum_{n \ge 0}R_n(u)v^n$ by directly taking derivatives of the 
above expression. But this would soon become undoable.
A concise way to calculate the coefficients is to make use of the
differential equation satisfied by $M(u,v)$. By extending the argument in
\cite{guttmann:1989} one arrives at
\begin{equation}
(1-u)[(1-v)^2u^2-2(1+6v+v^2)u+(1-v)^2]\frac{\partial}{\partial u}M(u,v)
+2v(1+u)M(u,v) = 0 .
\end{equation} 
This shows that the spontaneous magnetization is D-finite.
It is then determined implicitly from the recursion relation for $R_n$
\begin{eqnarray}
(n+3)(1-u)^2R_{n+3}(u)-[3(n+2)+2(5n+9)u+3(n+2)u^2]R_{n+2}(u) \nonumber\\
+[3(n+1)+2(5n+6)u+3(n+1)u^2]R_{n+1}(u) - n(1-u)^2R_n(u)  = 0.
\label{eq:recursion}
\end{eqnarray}

 From this recursion relation, it is clear that the partial sums $R_n$ have
poles only at $u = 1$. We should note that the recursion we have derived above
from D-finiteness is different from that discussed by Baxter\cite{baxter:1980}
and mentioned earlier.
In the latter case, to calculate the term $R_n$ one needs all $R_k$ prior to
$R_{n}$. By contrast, in the former case, only a fixed finite number of
terms is sufficient to obtain the term $R_n$ no matter how large $n$ is. 

 Examples of the case (ii) are the zero-field susceptibility of the 2D 
anisotropic Ising model and generating functions for self-avoiding polygons 
on the square and hexagonal lattices.
Guttmann and Enting argued using the finite-lattice method\cite{enting:1996} 
that infinitely many poles will appear in partial sums for those quantities 
in the limit $n \ra \infty$\cite{guttmann-enting:1996}.

 The few analyses mentioned above already indicate that Guttmann and Enting's 
conjecture unravels a new aspect of exactly solvable models, the connection of
the solvability to the distribution of poles and D-finiteness of the solution.
The conjecture deserves a further and detailed study. In particular, it is 
very important to examine whether Guttmann and Enting's conjecture has 
general applicability to a broad class of statistical mechanical systems. 
 The eight-vertex model is particularly suited to this end.
It can be regarded as a nontrivial generalization of the 2D Ising model. 
It has three coupling constants and can be compared with the previous analyses 
of the zero-field 3D Ising model and the 2D Ising model in a field.

 We adopt the spin-type model description of the eight-vertex model.
The spin variable $a$ placed on each site of the square lattice consisting of 
$N$ faces(Fig.1) takes the values $\pm 1$. The local Boltzmann weight is 
defined on each face $f$ in the lattice and depend on the spin configuration 
around the face 
$\{ \sigma \}_f = (a, b, c, d)$ as 
\begin{equation}
w( \{\sigma\}_f ) = R\exp(Kac+Lbd+Mabcd). 
\label{eq:2-1}
\end{equation}
The partition function is given by
\begin{equation}
Z_N(K,L,M) = \sum_{\sigma} \prod_f  w( \{\sigma\}_f ),
\label{eq:2-2}
\end{equation}
where the summation is over all spin configurations on the lattice sites and
the product is over all faces in the lattice. 
The lattice is assumed to have the periodic boundaries.
This model can be regarded as a pair of Ising models with anisotropic 
coupling constants $K$ and $L$, each living on one
of the sublattices and coupled by the four-spin coupling $M$.

 The local Boltzmann weight can be rearranged as
\begin{equation}
w( \{\sigma\}_f ) = \rho (1 + z_1 ac + z_2 bd + z_3 abcd),
\label{eq:2-3}
\end{equation}
where
\begin{eqnarray}
z_1 = \frac{t_1+t_2 t_3}{1+t_1 t_2 t_3},\rule{0.5cm}{0cm}
z_2 = \frac{t_2+t_3 t_1}{1+t_1 t_2 t_3},\rule{0.5cm}{0cm}
z_3 = \frac{t_3+t_1 t_2}{1+t_1 t_2 t_3},\\
\rho = R(\cosh K \cosh L \cosh M  + \sinh K \sinh L \sinh M),
\label{eq:2-4}
\end{eqnarray}
with $t_1 = \tanh K$, $t_2 = \tanh L$, $t_3 = \tanh M$.
$z_1$, $z_2$, $z_3$ are high-temperature variables and $\rho$ is the 
normalization factor.
 Let us represent each term in RHS of (\ref{eq:2-3}) graphically by
bonds as shown in Fig.2. We call the bonds shown in (b), (c), (d) of Fig.2
$z_1$-, $z_2$-, $z_3$-bonds.

Substituting (\ref{eq:2-3}) into (\ref{eq:2-2}) and expanding the products
yields all possible bond configurations over $N$ faces in the lattice.
The summation over spin configurations leaves only those in which an even 
number of bonds are incident to each lattice site, i.e., closed graphs.
 The reduced partition function, $\Lm_N(z_1,z_2,z_3) = (2\rho)^{-N} Z_N(K,L,M)$
, also plays the role of the generating function for the enumeration
of closed graphs in the lattice. They are classified in terms of the numbers
of the $z_1$-, $z_2$-, $z_3$-bonds, $l, m, n$. Then $\Lm_N$ can be written 
as
\begin{equation}
\Lm_N(z_1,z_2,z_3) = \displaystyle \sum_{l=1}^{\infty} \sum_{m=1}^{\infty}
		\sum_{n=0}^{\infty} P_{lmn}(N) z_1^{2l} z_2^{2m} z_3^{2n},
\label{eq:2-6}
\end{equation}
where $P_{lmn}(N)$ is the number of the closed graphs and is a polynomial in
$N$.
Define the reduced partition function per face in the thermodynamic limit by
$\Lm(z_1,z_2,z_3)=\displaystyle \lim_{N \ra \infty}(\Lm_N(z_1,z_2,z_3))^{1/N}$.
Then the expansion of $\Lm(z_1,z_2,z_3)$ in the connected closed graphs
is given by
\begin{equation}
\ln \Lm(z_1,z_2,z_3) = \displaystyle \sum_{l=1}^{\infty} \sum_{m=1}^{\infty}
		\sum_{n=0}^{\infty} a_{lmn} z_1^{2l} z_2^{2m} z_3^{2n},
\label{eq:2-7}
\end{equation}
where $a_{lmn}$ is the coefficient of $N$ in the polynomial $P_{lmn}$
\cite{domb:1974}.

 Our aim is to see whether or not $\ln \Lm(z_1,z_2,z_3)$ possesses 
the property conjectured by Guttmann and Enting. 
We begin by looking at the inversion relation for the eight-vertex model
\cite{baxter:1980}, which reads in the variables $z_1, z_2, z_3$ as
\begin{equation}
\ln \Lm(z_1,z_2,z_3) + \ln \Lm \Biggl(\biggl(\frac{1-z_2^2}{z_1^2-z_3^2}
\biggr)z_1, -z_2, -\biggl(\frac{1-z_2^2}{z_1^2-z_3^2}\biggr)z_3\Biggr) 
= \ln(1-z_2^2).
\label{eq:2-8}
\end{equation}
Note that if one sets the four-spin coupling $M$ to zero, i.e. $z_3 = z_1 z_2$,
(\ref{eq:2-8}) reduces to the inversion relation for the anisotropic Ising 
model. 
Since (\ref{eq:2-8}) involves the inversion of two variables, $z_1$ and $z_3$, 
one has to take the double infinite sum with respect to both to compare the
series (\ref{eq:2-7}) with (\ref{eq:2-8}).
 In practice this turns out to be a hard task.
Fortunately, one can circumvent the difficulty by a slight change of variables 
as $v_1 = z_1, v_2 = z_2$ and $v_3 = z_3/z_1$. Then (\ref{eq:2-8}) becomes
\begin{equation}
\ln \Lm(v_1,v_2,v_3) + \ln \Lm \Biggl(\biggl(\frac{1-v_2^2}{1-v_3^2}\biggr)
v_1^{-1}, -v_2, -v_3 \Biggr) = \ln(1-v_2^2).
\label{eq:2-9}
\end{equation}

 In this case, only $v_1$ is inverted and one has only to take the partial 
resummation with respect to $v_1$, which we write as
\begin{equation}
\ln \Lm(v_1,v_2,v_3) = \sum_{m=1}^{\infty} \sum_{n=0}^{\infty}
			 R_{mn}(v_1^2) v_2^{2m} v_3^{2n}.
\label{eq:2-10-a}
\end{equation}

 The corresponding expansion of the second term in LHS of (\ref{eq:2-9})
can be obtained by computing the Taylor expansion of 
$R_{mn}((1-v_2^2)^2/[v_1(1-v_3^2)]^2)$ with respect to $v_2^2$ and $v_3^2$. 
Collecting the terms containing the same powers in $v_2^2$ and $v_3^2$, one 
has the expansion,
\begin{equation}
\ln \Lm \Biggl(\biggl(\displaystyle
    \frac{1-v_2^2}{1-v_3^2}\biggr)v_1^{-1},-v_2,-v_3 \Biggr) 
    = \sum_{m=1}^{\infty} \sum_{n=0}^{\infty} Q_{mn}(v_1^2) v_2^{2m} v_3^{2n}.
\label{eq:2-10-b}
\end{equation}
The inversion relation (\ref{eq:2-9}) is now written as a perturbative 
relation,
\begin{equation}
R_{mn}(v_1^2) + Q_{mn}(v_1^2) = -\frac{1}{m} \delta_{n0}.
\label{eq:2-11}
\end{equation}
Note that $Q_{mn}$ depends only on $R_{kl}$ with $k+l$ no greater than $m+n$. 
The relation (\ref{eq:2-11}) gives constraints on the partial sums 
$R_{mn}(v_1^2)$ and is used in the later analysis of the PRSE.

 One might think that it is immediate to obtain the PRSE (\ref{eq:2-10-a})
for the eight-vertex model, since we have the exact solution at hand.
However, the expression of the exact solution in terms of the elliptic 
parameters is very implicit. So far we have not been able to derive the PRSE
from the exact solution and to find out the location of poles in the partial 
sums in $v_1^2$.

 Instead we employ a graphical approach to calculate $R_{mn}(v_1^2)$ in
the series expansion (\ref{eq:2-10-a}).
The closed graphs with $2m$ $z_2$-bonds, $2n$ $z_3$-bonds and an arbitrary
number of $z_1$-bonds contribute to $R_{mn}$.
Because of the presence of the four-spin coupling constant, there appear
vastly many types of closed graphs even at low orders. We have
developed a scheme to classify the graphs by introducing the notion of 
minimal graph, the graph with the least number of $z_1$-bonds among those
in one class. 
 The derivation of our method is lenghty and will be presented in a separate
paper. Here we report the results of our analysis of a first few terms of
the expansion:

 ${\bf m=1,n=0}$ \rule{2mm}{0mm}
The minimal graph is the square(Fig.3(a)). The remaining graphs are obtained by
successively adding a pair of $z_1$-bonds vertically, which we call 
$v_1^2$-extension. The weights of the $v_1^2$-extended graphs in (a) of Fig.3 
are $v_1^2 v_2^2$, $v_1^4 v_2^2$, $v_1^6 v_2^2$ and so on. 
Their contribution to $R_{10}$ sums up to $v_1^2/(1-v_1^2)$.

 ${\bf m=1,n=1}$ \rule{2mm}{0mm}
There are two classes of graphs with two $z_2$-bonds and
two $z_3$-bonds which contribute to $R_{11}$. 
One of them is shown in (b) of Fig.3.
Its minimal graph can be extended in three ways. It has two 
$v_1^2$-extensions, extending either of the two squares independently.
The third is a simultaneous extension of the two squares (see the third
graph of Fig.3 (b)), which we call $v_1^4$-extension.
The weights of the graphs in (b) of Fig.3 are thus
$v_1^4 v_2^2 v_3^2$, $v_1^6 v_2^2 v_3^2$, $v_1^8 v_2^2 v_3^2$ and so on. 
The contribution from these graphs sums up to $2v_1^4/(1-v_1^2)^2(1-v_1^4)$, 
where the factor $2$ takes account of embeddings of graphs and their
reflection.
The contribution from the other class to $R_{11}$ is given by 
$2v_1^6/(1-v_1^2)^2(1-v_1^4)$.

 ${\bf m=2,n=0}$ \rule{2mm}{0mm}
There are six classes of graph with four $z_2$-bonds and zero $z_3$-bonds 
which contribute to $R_{20}$. Five of the corresponding minimal graphs 
consist of connected graphs and the other of a disconnected graph with two 
squares. These graphs have only $v_1^2$-extensions.

 ${\bf m=1,n=2}$ \rule{2mm}{0mm}
There are twelve classes of graphs which contribute to $R_{12}$.
In each class the minimal graph can be extended in many different ways.
Some of the twelve minimal graphs are shown in Fig.3(c)-(e). 
The graphs of the class (c) are obtained from the minimal graph (c) by two
ways of $v_1^2$-extension and three ways of $v_1^4$-extension.
Their contributions yield $2v_1^6/(1-v^2)^2(1-v_1^4)^3$.
The contributions of the graphs of the class (d) and (e) can be found by
similar arguments: $4v_1^8/(1-v^2)^2(1-v_1^4)^3$ and 
$2v_1^{10}/(1-v^2)^2(1-v_1^4)^3$ respectively.
The sum of the contributions from the twelve classes turns out to take a
surprisingly simple form, $3v_1^6(1+v_1^2)/(1-v_1^2)^5$, which is free from
the pole at $v_1^2 = -1$.

 To summarize the above analysis, we have obtained the following expressions
for the first four terms of the PRSE:
\begin{eqnarray}
&R_{10}(v_1^2) = \displaystyle \frac{v_1^2}{1-v_1^2},
&R_{11}(v_1^2) = \displaystyle \frac{2v_1^4}{(1-v_1^2)^3},
\nonumber\\
&R_{20}(v_1^2) = \displaystyle \frac{v_1^2(2-5v_1^2+v_1^4)}{(1-v_1^2)^3},
&R_{12}(v_1^2) = \displaystyle \frac{3v_1^6(1+v_1^2)}{(1-v_1^2)^5}.
\label{eq:3-1} 
\end{eqnarray}
 The corresponding terms of $Q_{mn}(v_1^2)$ can be easily obtained from these
results as explained previously.

 There are a few important points to be noted. First, we have checked that
$R_{mn}(v_1^2)$ and $Q_{mn}(v_1^2)$ satisfy the perturbative form of the
inversion relation (\ref{eq:2-11}). This is a highly nontrivial check that
the result (\ref{eq:3-1}) is correct. The second point is concerned with 
the pole structure of $R_{mn}(v_1^2)$. 
As remarked above, $R_{10}(v_1^2)$ and $R_{20}(v_1^2)$ have a pole only at 
$v_1^2=1$ for a graphical reason.
As for $R_{11}(v_1^2)$, both of the contributions from the two classes
have poles at $v_1^2=1$ and $v_1^2=-1$. Curiously, the extra pole at
$v_1^2=-1$ cancels out in the sum of two contributions, leaving the pole term
$1/(1-v_1^2)^3$. It is even more miraculous that, as noted above, the same 
kind of cancellation of the poles at $v_1^2=-1$ takes place for 
$R_{12}(v_1^2)$, leaving the pole term $1/(1-v_1^2)^5$. 

 We think that the absence of extra poles at $v_1^2=-1$ in the first four terms
of the PRSE is not a pure accident but that it has a raison d'\^{e}tre.
It should be counted as a strong indication that poles in $R_{mn}(v_1^2)$ will
not appear at other values than $v_1^2=1$ for higher $m$ and $n$, supporting 
the conjecture of Guttmann and Enting.

 It is illuminating to compare the present result on the PRSE for the 
eight-vertex model with that for the 3D Ising model
\cite{hansel-et.al.:1987}.
The latter model also has three coupling constants. The inversion relation
and the PRSE in this model read as 
$\ln\Lm(t_1,t_2,t_3) + \ln\Lm(t_1^{-1},-t_2,-t_3)=\ln(1-t_2^2)$
and 
$\ln\Lm(t_1,t_2,t_3) = \displaystyle \sum_{mn}R_{mn}(t_1^2)t_2^{2m}t_3^{2n}$.
Nontrivial results start to appear from $R_{12}(t_1^2)$($R_{m0}(v_1^2)=
R_{0m}(v_1^2)$ 
coincide with $R_m(t_1^2)$ of the 2D Ising model, $R_{11}(t_1^2)$ has no poles
other than at $t_1^2=1$ for a straightforward graphical reason).
Indeed, $R_{12}(t_1^2)$ was found to have poles of the form 
$1/(1-t_1^2)^5(1+t_1^2)$. An extra pole at $t_1^2=-1$ already appears at the
third order ($m+n=3$) term of $R_{mn}(t_1^2)$. It is a sign that an infinite
number of poles will appear in the limit $m,n \ra \infty$. We note a clear
difference in the pole structure of $R_{mn}(t)$ between the eight-vertex model
and the 3D Ising model even at low orders. Probably this difference reflects 
the fact that the former model is solvable while the latter is not.

 Finally, we postulate the form of the partial sum $R_{mn}(v_1^2)$ at all 
orders by extending our results at low orders and examine the role of the
inversion relation regarding the exactly solvable partition function in the
approach of the PRSE.
 Let us assume that the partial sums of the eight-vertex reduced partition
function have only (multiple) poles at $v_1^2=1$. By a graphical consideration,
one can guess the general form of the partial sums:
$R_{mn}(v_1^2) = P_{mn}(v_1^2)/(1-v_1^2)^{2(m+n)-1}$,
where $P_{mn}(v_1^2)$ is a polynomial in $v_1^2$,
$P_{mn}(v_1^2) = \displaystyle \sum_{k=0}^{2(m+n)-1} c_{mnk} v_1^{2k}$.
Then $P_{mn}(v_1^2)$ can be determined recursively from the inversion and 
symmetry relations as follows. 
Suppose that $R_{kl}$ with $k+l < L-1$ are known. 
Then from the symmetry,
$\ln \Lm(v_1,v_2,v_3) = \ln \Lm(v_2,v_1,v_1v_3/v_2)$, we have the system of
linear equations,
\begin{equation}
\sum_{k=0}^{L-1}
\pmatrix{2(L-1)+r-k \cr 2(L-1) \cr} c_{mnk}
= \sum_{k=0}^{2r-1}
\pmatrix{2(r-1)+L-k \cr 2(r-1) \cr}
c_{r-n,n,k}, 
\label{eq:3-2}
\end{equation}
where $0 \le r \le L-1$ and $c_{mnk}=0 \mbox{ if } m<1$. 
Since the RHS of (\ref{eq:3-2}) is written with known coefficients only,
we have a half of the coefficients $\{ c_{mnk} : m+n=L, 0 \le k \le L-1\}$.
From the inversion relation (\ref{eq:2-9}) we have
\begin{equation}
R_{mn}(v_1^2) = -R_{mn}(v_1^{-2}) + \mbox{known terms},
\end{equation}
which determine the other half of the coefficients 
$\{c_{mnk}:m+n=L, L\le k\le 2L-1\}$.
Thus the solution for the eight-vertex partition function is reproduced 
by extending the method which was first used for the 2D Ising model 
by Baxter\cite{baxter:1980}.
The first few terms of this solution agree with (\ref{eq:3-1}).
Moreover, it reduces to the PRSE for the Onsager solution in the Ising limit.
This remarkable coincidence further supports our argument on the 
pole structure of $R_{mn}$ for the eight-vertex partition function.

 The authors thank Makoto Katori for bringing the paper of Guttmann and Enting
to their attention. They are indebted to Atsuo Kuniba for discussion at 
various stages of the present work, and to Satoru Odake for valuable comments
on the manuscript. They are also grateful to Tony Guttmann for elucidating
conversations, in particular on the differential equations to be obeyed by
solvable quantities.

\begin{figure}
\begin{center}
\epsfile{file=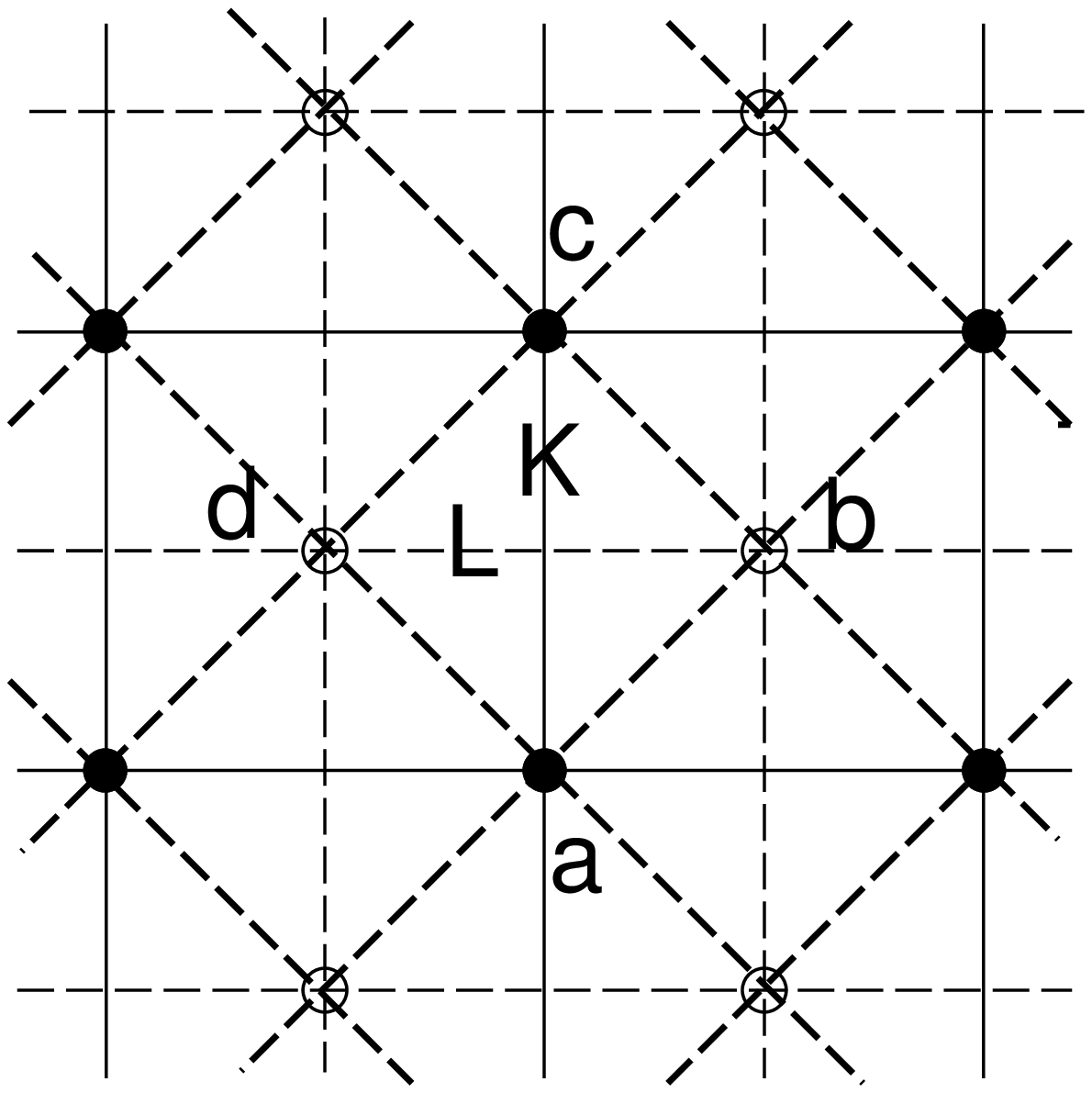,height=4cm}
\end{center}
\caption{The eight-vertex model.}
\end{figure}

\begin{figure}
\begin{center}
\epsfile{file=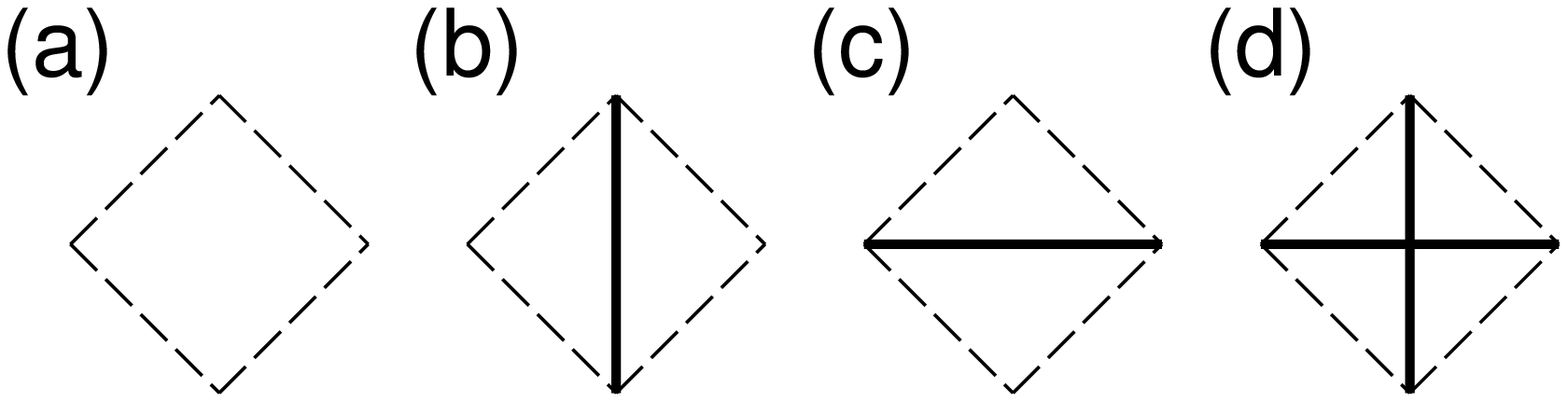,height=2cm}
\end{center}
\caption{Bond configurations on a face.}
\end{figure}

\begin{figure}
\begin{center}
\epsfile{file=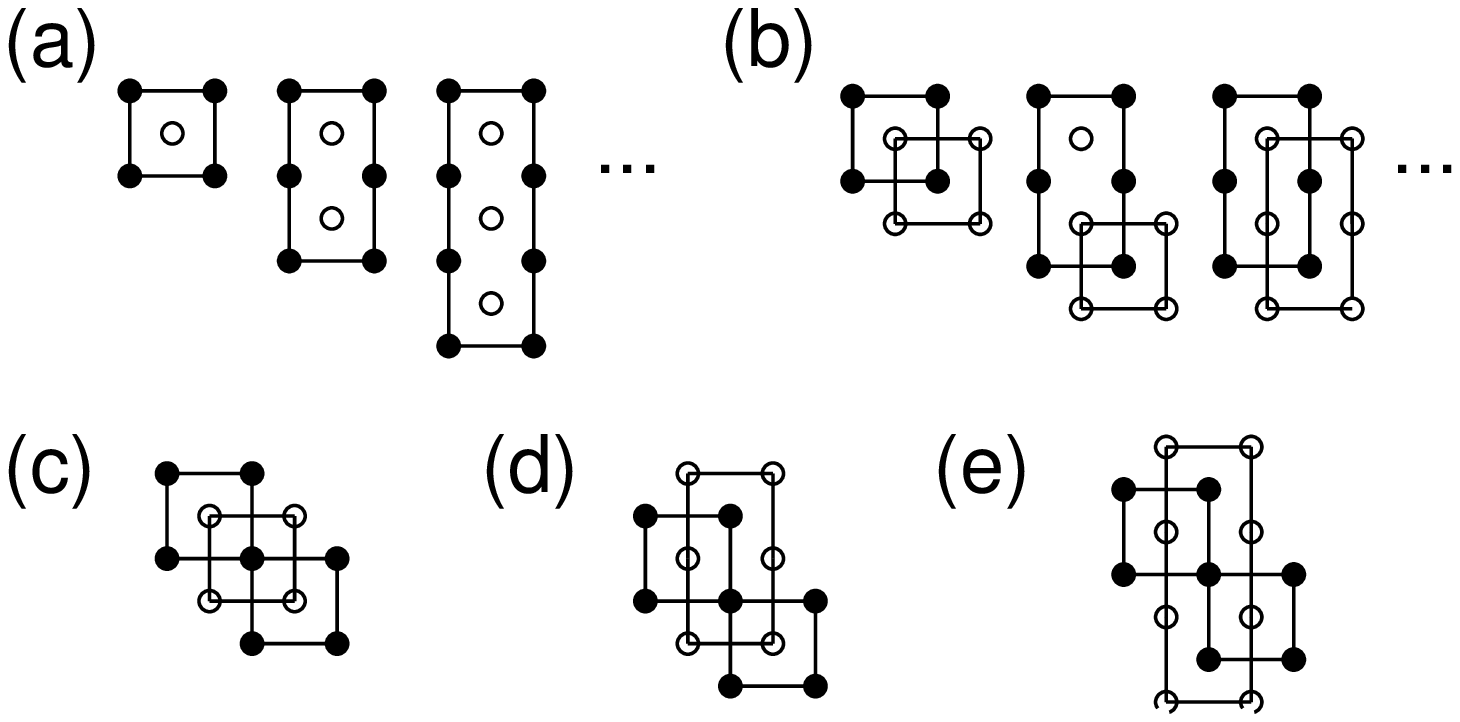,height=4cm}
\end{center}
\caption{Minimal graphs and their extensions.}
\end{figure}

\end{document}